\documentclass[preprint,12pt]{elsarticle}

\usepackage{graphicx}

\usepackage{amssymb}
\usepackage{amsmath}   

\journal{Journal of Computational Physics}

\begin{document}

\begin{frontmatter}



\title{Improved Maximum Entropy Analysis with an Extended Search Space}


\author{Alexander~Rothkopf}

\address{Fakult\"{a}t f\"{u}r Physik, Universit\"{a}t Bielefeld, 33615 Bielefeld, Germany}
\address{Albert Einstein Center for Fundamental Physics, Institute for Theoretical Physics, University of Bern, 3012 Bern, Switzerland}

\begin{abstract}
The standard implementation of the Maximum Entropy Method (MEM) follows Bryan \cite{springerlink:10.1007/BF02427376} and deploys a Singular Value Decomposition (SVD) to limit the dimensionality of the underlying solution space apriori. Here we present arguments based on the shape of the SVD basis functions and numerical evidence from a mock data analysis, which show that the correct Bayesian solution is not in general recovered with this approach. As a remedy we propose to extend the search basis systematically, which will eventually recover the full solution space and the correct solution. 
In order to adequately approach problems where an exponentially damped kernel is used, we provide an open-source implementation, using the C/C++ language that utilizes high precision arithmetic adjustable at run-time \cite{CODE}. The LBFGS algorithm is included in the code in order to attack problems without the need to resort to a particular search space restriction. 
\end{abstract}

\begin{keyword}

\MSC 62F15 \sep \MSC 49M15 \sep \MSC 90C53
\end{keyword}

\end{frontmatter}


\section{Introduction}

A wide variety of tasks in the physical sciences requires a deconvolution of raw data before relevant information can be accessed. Examples include the reconstruction of images of starlight having passed through a turbulent atmosphere or the extraction of spectral information from numerical simulations of the strong force. Here we are interested in the general setting, where the sought after and positive definite function, called spectrum $\rho(\omega)\geq 0$, is connected to data $D(\tau)$ via an integral kernel $K(\tau,\omega)$
\begin{align}
 D(\tau)=\int_{-\infty}^{\infty} K(\tau,\omega) \rho(\omega) d\omega.\label{Eq:Convol}
\end{align}
Depending on the choice of kernel function, Eq.\eqref{Eq:Convol} can amount to a Fourier-type transformation, where e.g. $K(\tau,\omega)\propto {\rm sin}[\omega\tau]$ or a double sided Laplace transform with $K(\tau,\omega)\propto {\rm exp}[-\omega\tau]$. In general the inversion of the above relation is an ill-defined problem and we will set out to give meaning to it through the use of Bayesian inference.

Let us start by preparing the stage, noting that data is obtained by an experimental apparatus or a numerical simulation and thus its values are known only at $N_\tau$ discrete points $D(\tau_i)=D_i$ in the interval $\tau_i\in[0,\beta]$, up to a given uncertainty denoted by an error matrix
\begin{align}
 C_{ij}=\frac{1}{N_{\rm c}(N_{\rm c}-1)}\sum_{k=1}^{N_{\rm c}} \Big(D^k_i- D_i \Big)\Big(D^k_j- D_j \Big).
\end{align}
Here $D^k_i$ represents one of the $N_{\rm c}$ individual measurements of the data-point at $\tau_i$.

To carry out the task of determining the spectrum from this data, we need to discretize $\rho(\omega_l)=\rho_l$ over frequencies $\omega_l$ in Eq.\eqref{Eq:Convol} using $N_\omega$ points between an upper and lower cutoff $\omega_{\rm max}$ and $\omega_{\rm min}$. This leads to a spacing of $\Delta\omega =\frac{\omega_{\rm max}-\omega_{\rm min}}{N_\omega}$.

Note that this step already requires us to supply additional knowledge about the measured system, since $\omega_{\rm max}$ and $\omega_{\rm min}$ need to be chosen such that all relevant frequencies encoded in the data can be accounted for. Prior information of this kind can often be derived from sampling theorems in the case of an experimental apparatus or the finite size of the underlying numerical simulation that produces the data-points.

Thus the fully discretized equation we are to supposed to invert reads
\begin{align}
 D_i=\Delta\omega \sum_{l=1}^{N_\omega} \;K_{il}\; \rho_l \label{Eq:ConvDiscr}
\end{align}

The task posed by inverting Eq.\eqref{Eq:ConvDiscr} is ill defined due to the presence of both noise in the measured data and the finite number of datapoints $N_\tau$, which is often significantly smaller than the number of points $N_\omega$ one wishes to reconstruct in the spectrum. 

Imagine performing a simple $\chi^2$ fitting, i.e. finding a set of points $\rho_l$ that reproduces the data $D_i$ within the errors $\sigma_i=\sqrt{C_{ii}}$. In such a case many degenerate solutions exist, none of which is superior to any other. The reason for this is that the finite number of data-points can only constrain parts of the spectrum. Unfortunately at this stage we are not able to decide which of the reconstructed features in $\rho$ these correspond to. Note that in addition, the problem at hand is not linear as might be assumed from Eq.\eqref{Eq:ConvDiscr}, since we require the values of $\rho_l$ to be positive definite. This in turn corresponds to an additional constraint to be met, which prohibits a naive matrix inversion in \eqref{Eq:ConvDiscr} even in the case of perfect data.

A possible way to give meaning to such a problem is provided by Bayesian inference. This well established branch of statistics tells us through Bayes theorem that prior information is a key ingredient to the question of what spectrum correctly describes the physical system under investigation. More precisely one asks, what is the probability of a test function $\rho_l$ to be the correct spectral function, given measured data $D_i$ and prior information $I$
\begin{align}
 P[\rho|D,I]=\frac{P[D|\rho]P[\rho|I]}{P[D|I]}.
\end{align}
The first term $P[D|\rho]$ appearing on the RHS is called the likelihood probability and denotes the probability of the data, given a test spectral function. This contribution is nothing but the usual $\chi^2$ fitting term and amounts to a Gaussian in the distance between measured data $D_i$ and the corresponding data $D^\rho_i$ obtained from inserting the test spectrum $\rho_l$ into Eq.\eqref{Eq:ConvDiscr}
\begin{align}
 P[D|\rho]\propto{\rm exp}[-{\cal L}]={\rm exp}\Big[ -\frac{1}{2} \sum_{i,j=1}^{N_\tau} (D_i-D^\rho_i)C_{ij}^{-1} (D_j-D^\rho_j) \Big].\label{Eq:LikelihodProb}
\end{align}
The second term on the RHS, the prior probability $P[\rho|I]$, is crucial in going beyond the naive $\chi^2$ fitting, as it incorporates our prior knowledge. We require the spectrum to be positive definite, hence this distribution may not permit negative values and we deploy the particular choice of the so called Shannon-Jaynes entropy ${\cal S}$ in the following
\begin{align}
 P_{MEM}[\rho|I(m)]\propto{\rm exp}[\alpha {\cal S}] = {\rm exp}\Big[ \alpha \sum_{l=1}^{N_\omega} \Big( \rho_l-m_l-\rho_l{\rm log}[\frac{\rho_l}{m_l}]\Big)\Big]. \label{Eq:PriorProb}
\end{align}
Here prior knowledge $I=I[(m)]$ is supplied through a function $m(\omega)$, which by definition denotes the correct spectrum in the absence of measured data. This function can e.g. contain the results of a previous investigation or an approximate solution obtained from theoretical considerations. Note that one has introduced a hyperparameter $\alpha$ in Eq.\eqref{Eq:PriorProb}, which is used to self consistently determine how strongly the entropy has to be weighted compared to the likelihood \cite{Jaynes1984_1,springerlink:10.1007/BF02427376,Jarrell1996133,Asakawa:2000tr}.

If we neglect the denominator $P[D|I]$, as it does not depend on the spectral function itself, the question of finding the most probable spectral function, given data and prior knowledge is now expressed as the following stationarity condition
\begin{align}
 \left. \frac{\delta}{\delta \rho_l} P[\rho|D,I(m)] \right|_{\rho=\rho_{\rm MEM}}\propto \left. \frac{\delta}{\delta \rho_l}\Big( P[D|\rho]P_{MEM}[\rho,I]\Big) \right|_{\rho=\rho_{\rm MEM}}=    0.\label{MEM:Optimize}
\end{align}
 Since the real exponential function is monotonous and we wish to avoid dealing with numbers over many orders of magnitude numerically, we focus in practice on the equivalent problem of minimizing the functional 
\begin{align}
  {\cal Q}(\rho,D,m) = {\cal L}(D,\rho)-\alpha {\cal S}(m,\rho).\label{Eq:Q}
\end{align}

To understand how the ill defined problem is given meaning, note that there are two contributions in Eq.\eqref{Eq:Q} that compete for the selection of the global minimum. Whereas ${\cal L}$ favors a spectrum that exactly reproduces the available datapoints, it is ${\cal S}$ that guides the spectrum toward the prior function. 

The most important fact to note is that there exists a proof \cite{Asakawa:2000tr}, which tells us the following. Since we supply in addition to our measured $N_\tau$ data-points $N_\omega$ points of prior information by introducing the function $m_l$, the functional ${\cal Q}(\rho,D,m)$ possesses a unique minimum in the $N_\omega$ dimensional space of functions $\rho_l$, if such an extremum exists\footnote{At this true global extremum, we expect the likelihood ${\cal L}$ to be of comparable size to the entropy term $\alpha{\cal S}$, all of them being of order ${\cal O}(1-10)$. If in the numerical implementation the most probable spectrum still remains at values of ${\cal L}$ larger than $\sim 100$ the discretization in frequency space is chosen too coarse or too narrow.}. This is not surprising, since with the inclusion of prior knowledge, we have at our disposal more points of data than free parameters entering the problem. Even the most extreme case, where no data is supplied, is well defined, as the 
prior function will then constitute the correct solution.

Intuitively the MEM result depends on a combination of three ingredients, the number of datapoints, the quality of the supplied data, as well as the prior information. The problem of inverting the underlying equation Eq.\eqref{Eq:ConvDiscr} is still ill-defined, but there exists a crucial difference to the naive $\chi^2$ fitting approach. Due to the presence of a prior function, Eq.\eqref{MEM:Optimize} selects a single solution from the degenerate set of functions that all equally well minimize $P[\rho|D]$. Part of this spectrum is fixed by the data points, part of it is selected through the function $m(\omega)$. I.e. changing the functional form of the prior will select a different spectrum, which however still reproduces the data within its errorbars. We can conclude that those parts of the spectrum that stay invariant under a change of prior must hence be fixed by the datapoints, while the rest of the spectrum follows from the choice of $m(\omega)$. 

How the recovered spectrum improves with increasing the number of datapoints or lowering the measurement errors depends in part on the form of the kernel function. In a Fourier-type setting, it is known that sampling the same interval $\tau\in[0,\beta]$ with an increasing number of points will allow us to reconstruct spectral features at higher and higher frequencies. Less errors on the other hand will allow us to improve the localization of peaks, i.e. the resolution of any individual peak will increase. In case of the Laplace transform, the number of sampled points is not connected to a maximum frequency but instead reflects in how reliably the width of a spectral peak can be recovered. 

\section{Towards an Improvement of the MEM Implementation}

In practice Eq.\eqref{MEM:Optimize} constitutes a high dimensional optimization problem, often of order $N_\omega\sim O(1000)$ and above. Since reliable second order methods, such as the Levenberg-Marquardt algorithm, require an inversion of the Hesse-matrix of size $N_\omega\times N_\omega$, this direct approach quickly becomes too costly when $N_\omega$ increases. One strategy, which was introduced in \cite{springerlink:10.1007/BF02427376} is to limit the dimensionality of the solution space apriori by choosing a set of basis functions derived from an SVD of the discretized integral kernel $K^t_{il}$.  The apparent reduction of computational cost in this approach is significant, it posits that one has to deal only with $N_\tau$ degrees of freedom instead of the original $N_\omega$.

We will show in the following that the solution from within the SVD search space does not in general correspond to the global minimum sought after in Eq.\eqref{MEM:Optimize}. Our argument is based on the functional form of the basis functions following from the SVD of the kernel on the one hand and a direct counterexample from a mock data analysis, which shows how Bryan's method fails to obtain the correct Bayesian solution.

Before elaborating on a possible improvement let us briefly recollect how the standard implementation is justified.

\subsection{Bryan's Search Space}

Inserting the definitions of Eq.\eqref{Eq:LikelihodProb} and Eq.\eqref{Eq:PriorProb} into the stationarity condition for the functional ${\cal Q}$
\begin{align}
 \frac{\delta {\cal Q}(\rho,D,m)}{\delta \rho}=0,
\end{align}
we find the following implicit expression for the spectrum
\begin{align}
 -\alpha {\rm log}[\frac{\rho_l}{m_l}]=\sum_{i=1}^{N_\tau} K_{il} \frac{d{\cal L}}{dD^\rho_i(\rho)},
\end{align}
the LHS of which  originates from the entropy term. The fraction in the logarithm invites us to make the positive definiteness of the spectrum and the prior function explicit by using the general parametrization $\rho_l=m_l\, {\rm exp}[a_l]$, which, if written in vector notation, leads to
\begin{align}
 -\alpha \vec{a} =  K^t \vec{\frac{d{\cal L}}{dD^\rho(a)}}.
\end{align}
Note that $\vec{a}$ essentially characterizes the deviation of the spectrum from the prior function. Bryan's strategy amounts to applying the SVD to the transposed kernel $K^t=U\Sigma V^t$, such that 
\begin{align}
 -\alpha \vec{a} = U \Sigma V^t  \vec{\frac{d{\cal L}}{dD^\rho(a)}}. \label{Eq:DefSVDsp}
\end{align}
Note that by definition of the SVD, the matrix $U$ contains a full orthonormal basis of the $\mathbb{R}^{N_\omega}$. $\Sigma$ on the other hand is a diagonal matrix, which contains only $N_\tau$ entries different from zero, since there were only $N_\tau$ columns in $K^t_{il}$. The above implicit expression leads Bryan to the incorrect (as will be shown in the next section) assumption that the vector $\vec{a}$, characterizing the global extremum, always has to lie in the subspace spanned by the first $N_\tau$ columns of the matrix $U$. He thus decides to parametrize the spectral function using the $N_\tau$ values $b_j$ 
\begin{align}
 \rho_l=m_l \,{\rm exp}[\sum_{j=1}^{N_\tau} U_{lj} b_j].\label{Eq:BryanParam}
\end{align}

\subsection{Inadequacy of the search space}

The first sign of an inadequacy of the search space introduced through the parametrization in Eq.\eqref{Eq:BryanParam} can be found in the functional form of the basis functions $U_{lj}$. 

\subsubsection{SVD Basis functions}

\begin{figure*}[!t]
\hspace{-1.3cm}
 \includegraphics[scale=0.2, angle=-90] {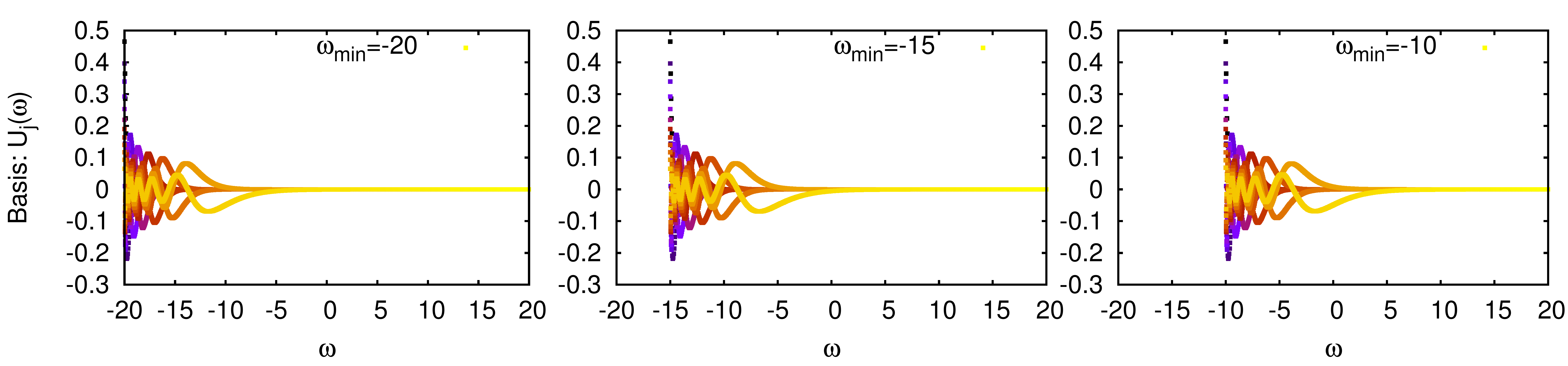}
 \caption{ Comparison of the first twelve basis functions $U_j(\omega)$ from an SVD of the kernel $K(\tau,\omega)={\rm exp}[-\omega\tau]$. For the discretization we choose $N_\tau=12$ with $\tau\in[0,6.1]$, while the frequency interval with upper cutoff $\omega_{\rm max}=20$ uses a spacing of $\Delta\omega=0.02$. From the left to the right panel we change the lower cutoff of the $\omega$ range $\omega_{\rm min}=-20,-15,-10$ and observe that the functional form of the $U_j(\omega)$'s does not change, while they are shifted as a whole along the frequency axis. Note that already for the choice $\omega_{\rm min}=-10$ the oscillatory regime ends slightly above $\omega_{\rm osc}\simeq0$}
\label{Fig:SVDBasisShift}
\end{figure*}

In Fig.\ref{Fig:SVDBasisShift} we plot the first twelve basis functions for the case of the Laplace transform with $K(\tau,\omega)=e^{-\omega\tau}$. The frequencies are discretized with a $\Delta\omega=0.02$ in three different intervals, ranging from a common upper cutoff $\omega_{\rm max}=20$ to $\omega_{\rm min}=-10,-15,-20$. What we find is that all functions $U_j(\omega)$ share the same qualitative behavior. Starting from $\omega_{\rm min}$ they oscillate up to a certain $\omega_{\rm osc}$, beyond which a rapid damping toward zero sets in. If we choose (Fig.\ref{Fig:SVDBasisShift}, right) $\omega_{\rm min}=-10$, while being fixed to $N_\tau=12$ basis functions, the oscillatory part extends only up to $\omega<\omega_{\rm osc}\simeq0$. Obviously we will not be able to reconstruct sharp peak structures in the region $\omega>\omega_{\rm osc}$. 

This constitutes a conceptual problem in the approach of Bryan, since the derivation of Eq.\eqref{Eq:BryanParam} did not refer to a particular choice of $\omega_{\rm min}$ and thus allows us to set its value arbitrarily.  As seen from the center and left panels in Fig.\ref{Fig:SVDBasisShift}, changing $\omega_{\rm min}$ while keeping $\Delta\omega$ fixed, does not influence the length of the oscillatory regime but only shifts the whole function to lower frequencies. It is thus possible to always make the MEM fail within the singular search space, since $\omega_{\rm min}$ can be large and negative, such that no peak structures remain available for a reconstruction of the spectrum. 

Note that the proof of existence and uniqueness for the solution laid out in \cite{Asakawa:2000tr} does not rely on any parametrization or restriction of the underlying functional space. The fact that by choosing $\omega_{\rm min}$, Bryan's MEM can always be made to fail, indicates that the $N_\tau$ dimensional subspace artificially restricts the solution of Eq.\eqref{MEM:Optimize}.

The effects of Bryan's search space on the quality of a reconstruction of actual spectra can be investigated by using mock data, as we will proceed to do in the next section  

\subsubsection{Numerical Evidence from a Mock Data Analysis}

Working with data from numerical simulations of the strong force \cite{Rothkopf:2011db}, it became apparent that the MEM based on Bryan's prescription was not able to adequately reconstruct the encoded spectrum in many cases. Here we demonstrate this effect by feeding to the algorithm a set of prepared datapoints, which encode a known spectral function, whose form closely resembles those encountered in our numerical investigation. 

The spectrum used in the following is a particular choice, it however contains several elements that are characteristic for those cases where Bryan's approach warrants an improvement. If we e.g. had only a single peak encoded in the spectrum, we might be able to improve the situation somewhat by moving $\omega_{\rm min}$ close to expected position of that spectral feature. In nature however we often encounter the case that several peaks of different width and wildly different amplitude are distributed over a broad frequency range, hence the adjustment of $\omega_{\rm min}$ is not an adequate remedy. Therefore we choose as mock spectrum a sum of four Gaussian peaks with parameters as shown in Tab.\ref{Tab:GPeaks}.

\begin{table}[!t]
\begin{center}
\begin{tabular}{ l || c| c | c | r  }
  &1st peak & 2nd peak & 3rd peak & 4th peak\\
  \hline  \hline                       
  amplitude:& $3e^{-8}$ & 0.6 & 0.25 & 0.2 \\
  position:& -2.3 & 0.52 & 2.6 & 7.5 \\
  width: & 0.1 & 0.1 & 0.4 & 1.4 \\
\end{tabular}
\caption{Parameters of the Gaussian peaks used in the mock function $\rho_{\rm mock}$, inspired by Lattice QCD data obtained in \cite{Rothkopf:2011db} }
\label{Tab:GPeaks}
\end{center}
\end{table}

The frequency range of $\omega^{\rm mock}\in[-5,20]$ is discretized with $N^{\rm mock}_\omega=5000$ points used to sample the mock spectrum and to generate ideal data $D^{\rm ideal}$ through insertion into Eq.\eqref{Eq:ConvDiscr}. The influence of errors is taken into account by adding Gaussian noise at each individual $\tau_k$ with variance $\delta D^{\rm mock}_k$. The strength of the disturbance is controlled by the parameter $\eta$, i.e.

\begin{align}
 \delta D^{\rm mock}_k=k\eta D^{\rm ideal}_k, \quad k\in[1,\cdots,N_\tau].
\end{align}

As we wish to separate the question of how well the reconstruction succeeds from the quality of data and focus on the choice of search space, a small noise $\eta=0.0001$ is used to only slightly distort the ideal mock data. 

We choose as prior the function 
\begin{align}
 m(\omega)=\frac{1}{\omega+\omega_0},\label{Eq:mfunc}
\end{align}
with $\omega_0$ selected such that its integral coincides with the area under the mock spectrum. Any particular choice of the prior will influence the outcome of the reconstruction, since the parametrization of Eq.\eqref{Eq:BryanParam} includes $m(\omega)$ as a prefactor\footnote{As we argued in the introduction, the result of the MEM will be a spectrum, parts of which are constrained by the data, parts of which are constrained by our choice of $m(\omega)$. If our goal is to reliably determine, which part of $\rho(\omega)$ is actually a result of the supplied measurements, we will have to redo the MEM with several different priors to identify, what spectral feature remains unchanged.}. In practice we often only have partial prior information available, usually far from the region where the spectral features of interest are located. Hence our goal here is to use a prior that resembles this fact, by approaching zero for large frequencies, while being incorrect but still a smooth function at small frequencies.

To reconstruct the supplied mock spectrum, we choose for the MEM the frequency range $\omega\in[-10,20]$ divided into $N_\omega=1500$ points, whereas $\tau\in[0,6.1]$ with $N_\tau=12$. The inclusion of negative frequencies leads to a large dynamic range of the kernel, hence the internal arithmetic is set to use 384 bits of precision.

\begin{figure*}[t!]
\hspace{-1.3cm}
\includegraphics[scale=0.22,angle=-90] {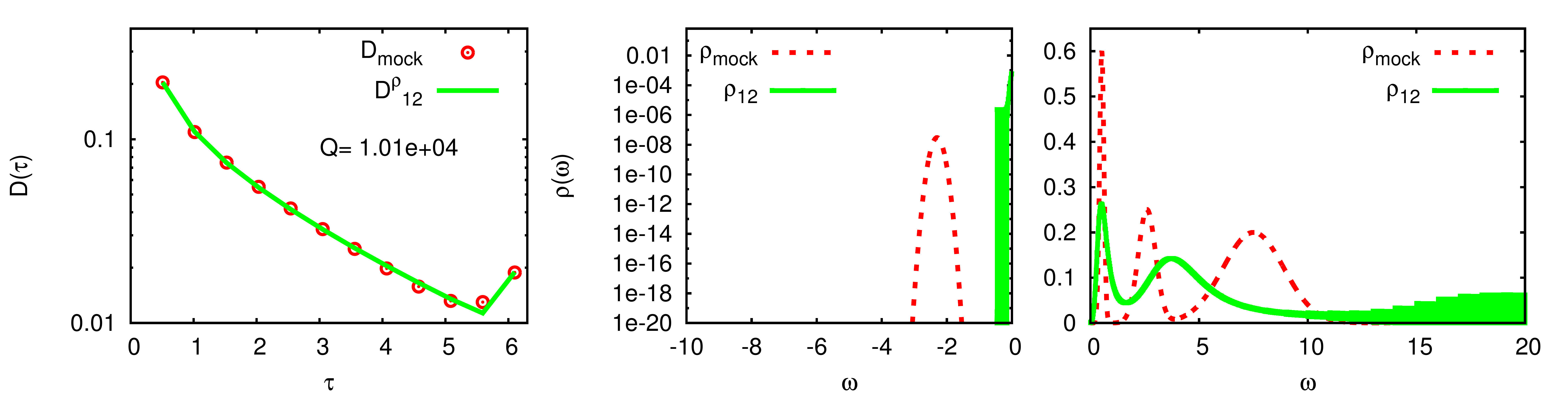}
 \caption{(left) Comparison of the $N_\tau=12$ mock data points (circles) and the data (line) obtained from inserting the MEM reconstructed spectrum into Eq.\eqref{Eq:ConvDiscr}. Note that with Bryan's prescription used here, the solution does not reproduce the datapoints around $\tau\simeq5$ within their errorbars. (center and right) Comparison of the mock spectrum and the reconstructed function $\rho(\omega)$ according to Bryan's prescription. Note that the peak at negative frequencies is not captured at all, as is the third peak at positive $\omega$.}
 \label{Fig:BryanMockReconstr}
\end{figure*}

In Fig.\ref{Fig:BryanMockReconstr} we present the results of the reconstruction according to Bryan's prescription. The first indication that the MEM has not been successful in this approach is the large value of the residual ${\cal Q}\simeq 10000$, which is dominated by a large value of ${\cal L}$ of the same order of magnitude. Indeed the idea of the MEM is to regularize an otherwise underdetermined $\chi^2$ fitting, by selecting from a large number of degenerate solutions the one with maximum entropy. This however entails that the chosen solution still reproduces all data within their errors, which is only possible if ${\cal L}\sim {\cal O}(1)$.

Looking at the reconstructed spectrum itself in the center and left plot of Fig.\ref{Fig:BryanMockReconstr} we find that the negative frequency peak as well as the third peak at large $\omega$ is not captured at all, while the first two peaks at $\omega>0$ are washed out and shifted. This is not surprising if we remember the set of basis functions available to the MEM in this case, as shown in the right panel of Fig.\ref{Fig:SVDBasisShift}. Within Bryan's approach their number is fixed by the quantity of available data-points. In addition, our choice of $\omega_{\rm min}=-10$ is valid, as we expect from the upward trend in the mock data that negative frequencies need to be taken into account. Since the oscillating range of the functions $U_j(\omega)$ ends shortly above $\omega=0$ it is however very difficult to reproduce the correct spectral features. 

We conclude that the search space provided by the first $N_\tau$ columns of the SVD of the transposed kernel $K^t_{il}$ does not allow us to reconstruct reliably the spectrum encoded in the mock data $D^{\rm mock}$. Thus we set out to improve the implementation of the maximum entropy method by extending the search space systematically as laid out in the following section.

\subsection{Extension of the search space}

The reason for the popularity of Bryan's approach is that it apparently offers a dramatic decrease in computational cost from $N_\omega$ to $N_\tau$ degrees of freedom. However the proof on the existence and uniqueness of an MEM solution in \cite{Asakawa:2000tr} applies only to the full $\mathbb{R}^{N_\omega}$ search space. In addition we have seen that the reconstruction in the SVD subspace can always be made to fail by choosing $\omega_{\rm min}$ large and negative. 

Therefore we propose to systematically enlarge the search space starting from Bryan's SVD subspace with the prospect of locating the correct global extremum of the functional ${\cal Q}(\rho,D,m)$ already with a number $N_{\rm ext}<N_\omega$ of basis functions. To this end we decide to extend the search space by including more and more of the columns of the matrix $U$ in the parametrization of the spectrum, so that now
\begin{align}
 \rho_l=m_l \,{\rm exp}[\sum_{j=1}^{N_{\rm ext}} U_{lj} b_j ]\label{Eq:MeParam}
\end{align}
with $N_\tau<N_{\rm ext}<N_\omega$. 

The number of basis vectors required to adequately determine the global extremum can then be determined by increasing the number $N_{\rm ext}$ until the minimal value of ${\cal Q}(\rho,D,m)$ does not decrease when adding an additional basis function. In the worst case this process has to be continued until $N_{\rm ext}=N_\omega$ since only the full set of columns of $U$ encodes a complete set of basis vectors for the $\mathbb{R}^{N_\omega}$.

\begin{figure*}[!t]
\hspace{-1.5cm}
 \includegraphics[scale=0.22,angle=-90] {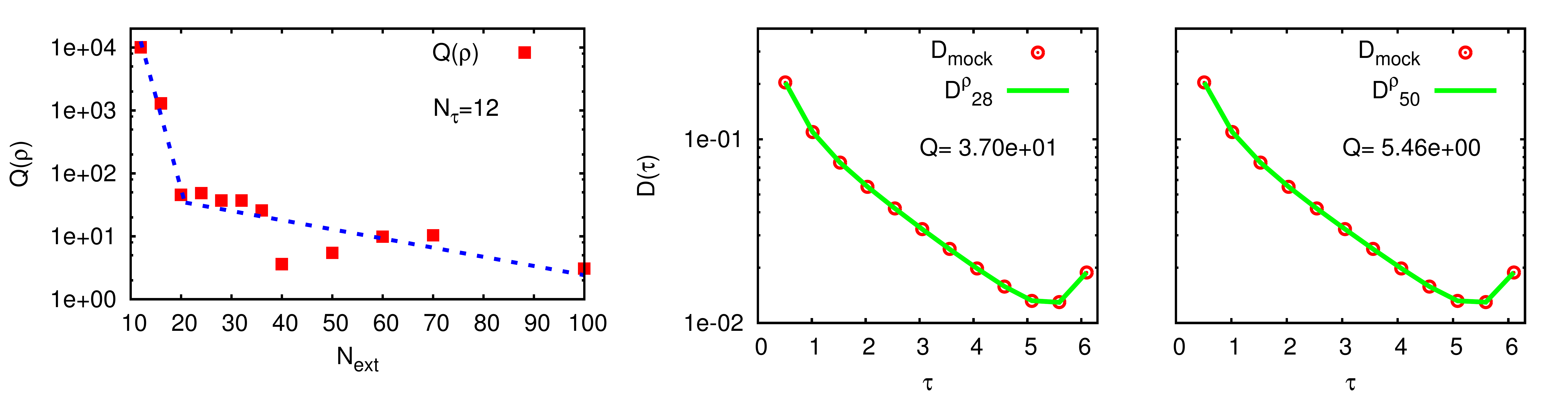}
  \caption{(left) The values of ${\cal Q}$ associated with the final MEM reconstruction for different numbers of basis vectors used in the parametrization Eq.\eqref{Eq:MeParam}. Note that all runs use the same $N_\tau=12$ mock dataset so that the difference in the value of ${\cal Q}$ solely originates in the available search space. This result is a direct counterexample to the claim that the correct MEM solution, i.e. the global extremum of Eq.\eqref{MEM:Optimize} always lies in Bryan's SVD search space. (right) Comparison of the mock data (circles) and the values (line) obtained from inserting the MEM reconstructed spectrum for $N_{\rm ext}=28$ and $50$ into Eq.\eqref{Eq:ConvDiscr}. The large discrepancy at $\tau\simeq5$ that existed in the case $N_{\rm ext}=N_{\tau}$ is significantly reduced here. }
\label{Fig:QvalExtended}
\end{figure*}

The central result, concerning the increase in the number of basis vectors, can be found in the right panel of Fig.\ref{Fig:QvalExtended}. There we plot the dependence on $N_{\rm ext}$ of the value of ${\cal Q}$, associated with the final solution of the MEM reconstruction. Contrary to the claim of Bryan, the global minimum sought after in Eq.\eqref{MEM:Optimize} is found outside of the SVD search space. Instead, after a rapid decrease of the residual ${\cal Q}$ for $12<N_{\rm ext}<20$ the reconstruction further improves at a slower rate and we are able to reach the region of ${\cal Q}\sim O(1-10)$ in which the correct solution is supposed to be located. The decrease in ${\cal Q}$ is also directly related to the success in reconstructing the mock data shown on the right of Fig.\ref{Fig:QvalExtended} for the values $N_{\rm ext}=28$ and $50$. While in the case of Bryan's search space with $N_{\rm ext}=N_\tau$, shown in the right panel of Fig.\ref{Fig:BryanMockReconstr}, the data at $\tau\simeq5$ was not 
reproduced within its errorbars, the discrepancy is significantly reduced here.

\begin{figure*}[!t]
\hspace{-1.5cm}
\centering
 \includegraphics[scale=0.27,angle=-90] {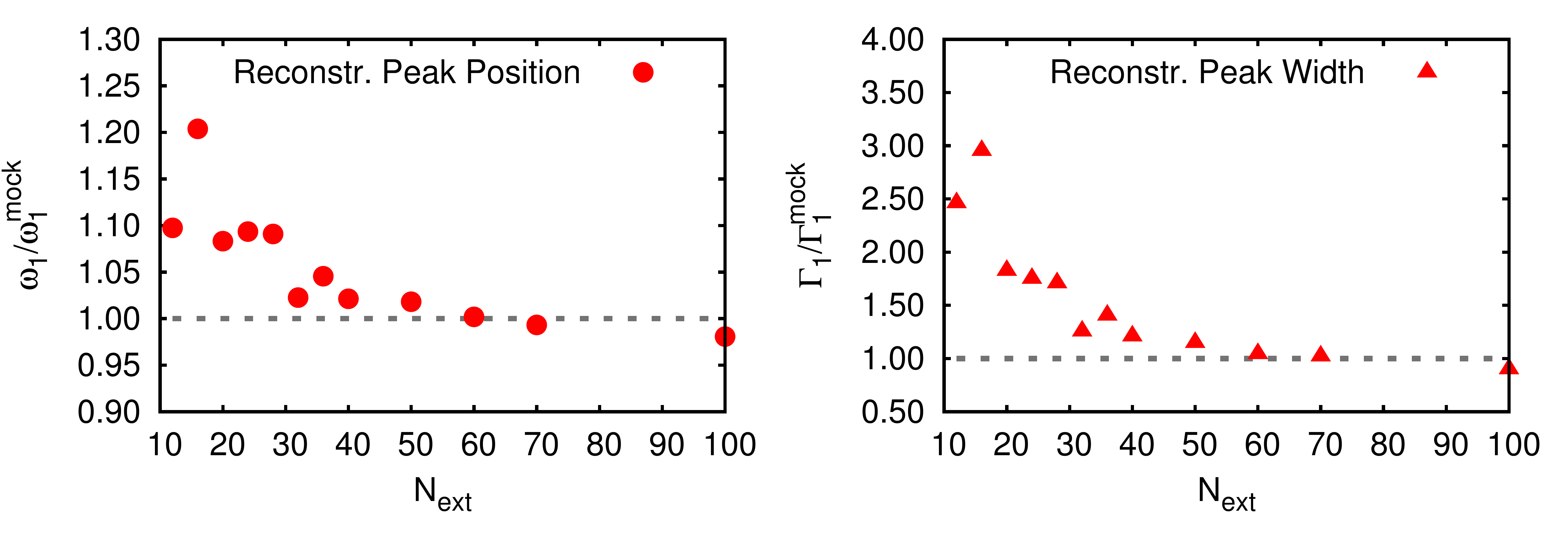}
  \caption{Visualization of the improvement in reconstructing the spectrum through an increase in the number of basis functions $N_{\rm ext}$. We plot the relative deviation of the reconstructed peak position $\omega_1/\omega_1^{\rm mock}$ against the number of supplied basis functions on the left. The right panel on the other hand shows the relative deviation of the reconstructed peak width  $\Gamma_1/\Gamma_1^{\rm mock}$. }
\label{Fig:ExtendedreconstrParameters}
\end{figure*}

Alternatively we can also observe an improvement in the recovery of the mock spectrum parameters. As an example we fit the lowest lying positive peak of the MEM result and compare the extracted values to the mock parameters of Tab.\ref{Tab:GPeaks}. Fig.\ref{Fig:ExtendedreconstrParameters}, which shows the relative deviation of the extracted parameters, tells us that both the reconstruction of the peak position and width improves as we increase the value of $N_{\rm ext}>N_{\tau}$.  For small values of $N_{\rm ext}$ the MEM tends to overestimate the position of the peak, since it tries to incorporate the higher lying spectral features into the insufficient number of degrees of freedom available to it. The width is also initially estimated with a too large value, since the oscillatory behavior of the basis functions is not fast enough to reproduce a narrow structure as small as the first peak\footnote{Note that for larger values of $N_{\rm ext}>60$ both the width and position in Fig.\ref{Fig:ExtendedreconstrParameters} are being underestimated, as the basis functions are able to produce structures with a width smaller than the lowest lying peak. This issue can be remedied if a larger number of data-points is supplied.}.

\begin{figure*}[!t]
\hspace{-1.2cm}
 \includegraphics[scale=0.20,angle=-90] {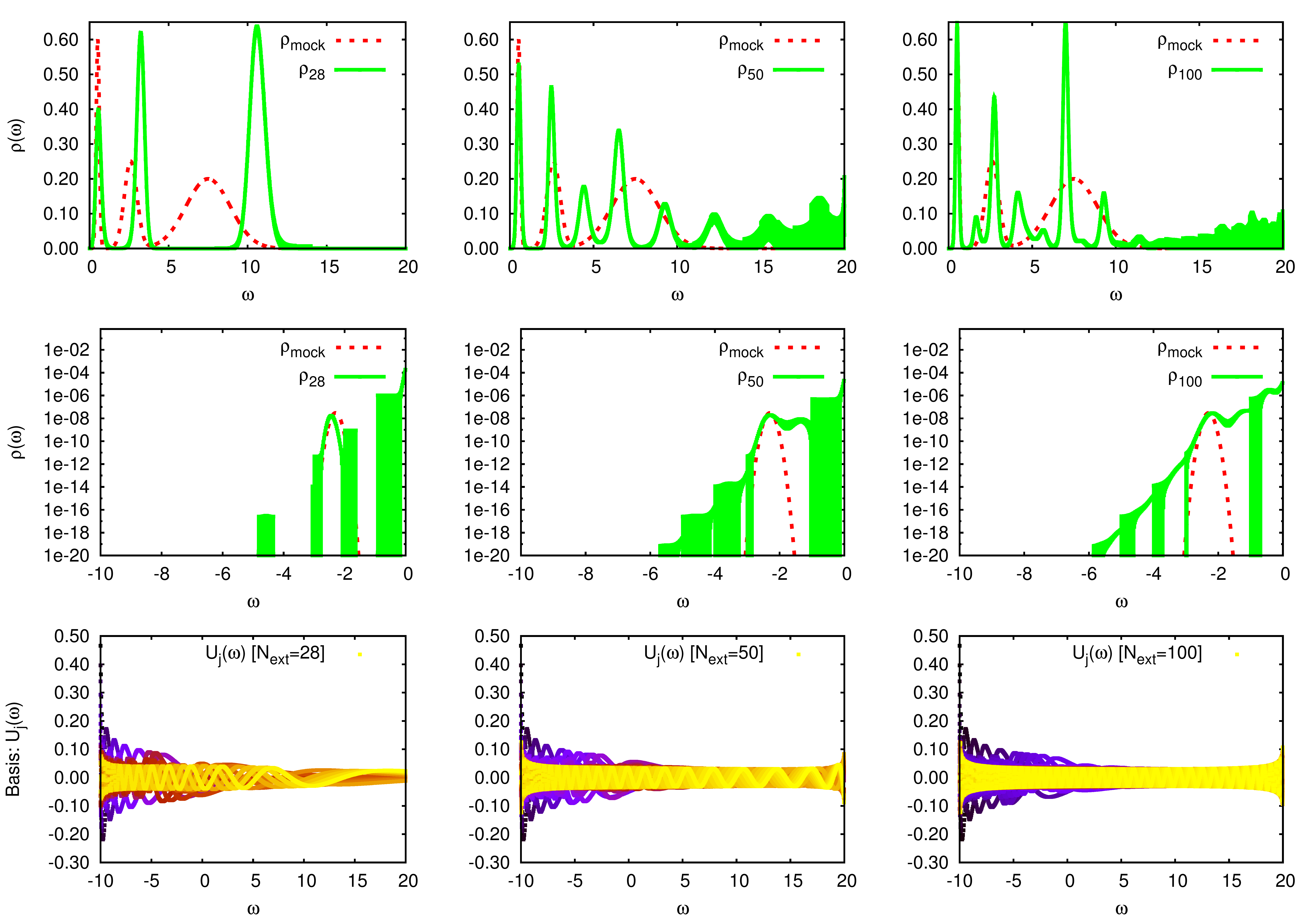}
  \caption{Comparison of the reconstructed spectra along the positive frequencies (top row) and negative frequencies (middle row). As the lowest lying positive peak is better and better reconstructed when going from $N_{\rm ext}=28$ (left column) via $N_{\rm ext}=50$ (center column) to $N_{\rm ext}=100$ (right column) it is clearly visible that at higher frequencies lots of wiggly structures arise. As argued in the text, the data-points are only able to constrain parts of the spectrum, the rest being determined by our choice of $m(\omega)$. To identify which of the wiggly features are actually constrained by the supplied measurements, we need to redo the MEM with a different functional form of the prior and observe their variation. (bottom row) The set of basis functions used in the determination of the MEM spectrum. }
\label{Fig:ExamplesExtendedReconstr}
\end{figure*}

In order to inspect the overall changes in the reconstruction of the mock spectrum brought about by an extension of the search space, we provide Fig.\ref{Fig:ExamplesExtendedReconstr}. There we plot the full spectrum at positive (top row) and negative frequencies (middle row) as well as the available basis functions (bottom row) for three different values of $N_{\rm ext}=28,50$ and $100$ (left, center and right column). While we find that in accordance with Fig.\ref{Fig:ExtendedreconstrParameters} the lowest lying positive frequency peak is increasingly well captured, the higher omega region shows a marked increase in variation. To understand which of these spectral features are actually important to us, we need to remember the role of the prior function. The result of the MEM reconstruction depends both on the supplied data and the choice of $m(\omega)$. As part of the spectrum is fixed by the former, part of it by the latter, we need to redo the MEM with different functional forms for the prior and observe which region stays invariant, subsequently being identified as constrained by the data.

\begin{figure*}[!t]
\hspace{-1.5cm}
\centering
 \includegraphics[scale=0.25,angle=-90] {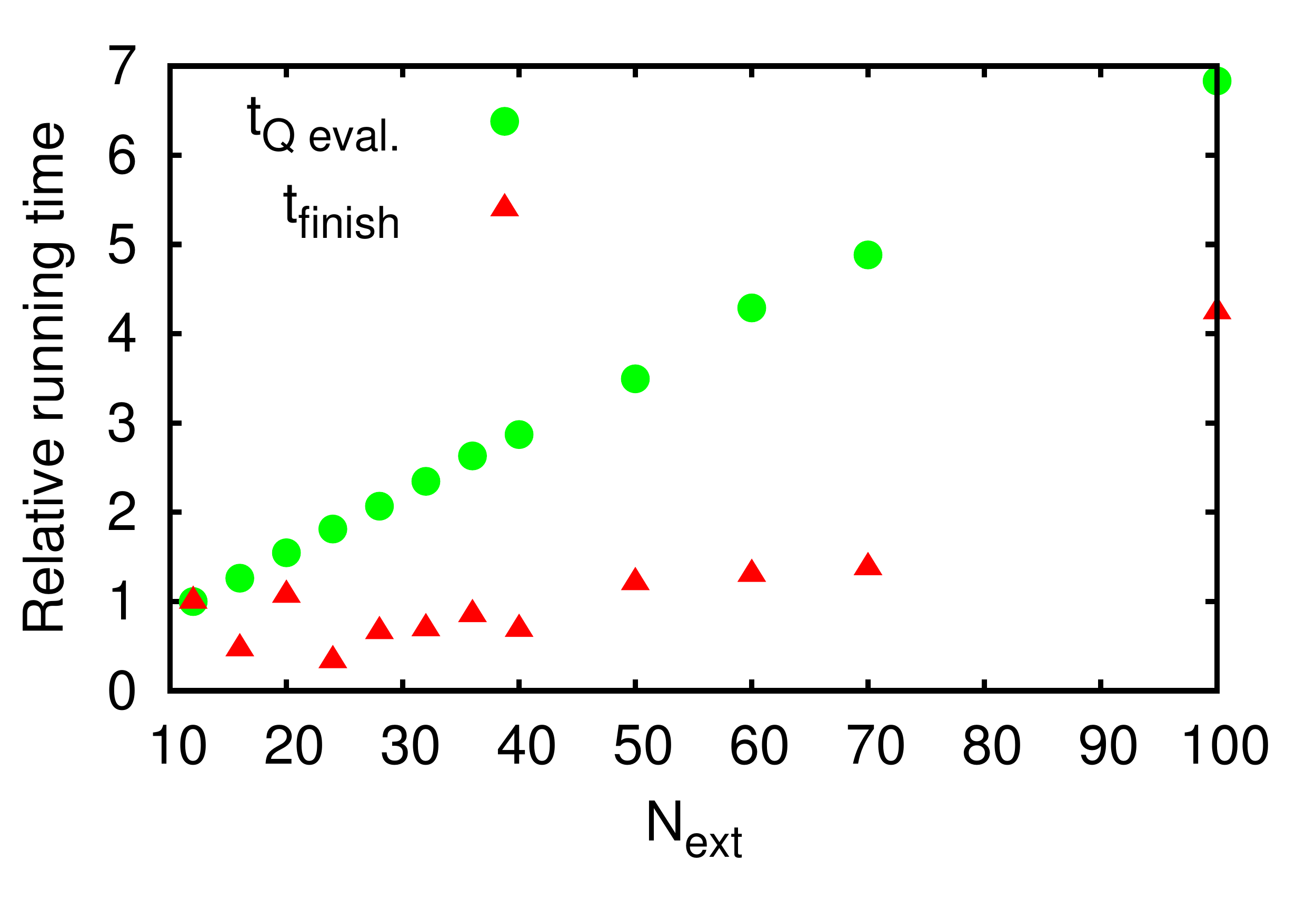}
  \caption{Comparison of running time for the evaluation of the functional ${\cal Q}$ (circle) and the overall running time of the program (triangle) relative to the values at $N_{ext}=N_{\tau}=12$. As expected the individual function evaluation time grows linearly with $N_{\rm ext}$ as only a linear increase of additions contributes to Eq.\eqref{Eq:MeParam}. The overall running time also shows a slowing down for larger values of $N_{\rm ext}$, however the  behavior for small numbers of basis functions does not exhibit a clear trend. A possible explanation is that for small $N_{\rm ext}$ the search space is too limited to approach the vicinity of the correct extremum, hence the minimizer will use a lot of time along the boundary of the restricted search space before settling into a local minimum.}
\label{Fig:SearchTime}
\end{figure*}

We have seen that by increasing the number of basis functions the quality of the reconstructed MEM spectrum can be significantly improved. The price to pay is an associated increase in computational cost. The most direct consequence of a larger number of basis vectors is that the evaluation time of the function ${\cal Q}$ increases linearly with $N_{\rm ext}$ as expected from Eq.\eqref{Eq:MeParam} and confirmed by explicit timing in Fig.\ref{Fig:SearchTime} (circles). The overall running time of the program increases monotonously (triangles), once $N_{\rm ext}>40$ but for smaller values the required time varies strongly. The reason is that the minimizer in the case of a severely restricted set of basis functions will only be able to move into the direction of the global minimum until it reaches the boundary of the search space, where it remains for a long time before settling into a local minimum. 

\section{Conclusion}

The Maximum Entropy method offers a solution to the question of how to bring meaning to the ill-defined problem of inverting Eq.\eqref{Eq:ConvDiscr}, i.e. to infer the $N_\omega$ values $\rho_l$ from a noisy and finite data-set $D_i$ of size $N_\tau$. Instead of maximizing only the likelihood probability with a test spectral function $\rho_l$, one regularizes the process by including as prior probability the Shannon-Jaynes entropy. The function $\rho^{\rm MEM}_l$ that represents the extremum of Eq.\eqref{MEM:Optimize} is hence the most probable answer in the Bayesian sense.

Since in Bryan's approach the selection of the SVD basis functions does not depend on the choice of $\omega_{\rm min}$ and their number is fixed by the supplied number of data-points, we argue that his search space does not in general contain the correct global extremum of the functional ${\cal Q}(\rho,D,m)$. Numerical evidence was presented to support this conclusion. We thus propose to systematically expand the search space to $N_\tau<N_{\rm ext} <N_\omega$ dimensions until the correct global extremum of the functional has been found.

Introducing a large number of basis functions inevitably leads to the appearance of ''wiggly`` structures in the reconstructed spectral function $\rho^{\rm MEM}(\omega)$. If they are not constrained by the data, such artifacts can be identified through a variation of the prior function. In turn, the features of $\rho^{\rm MEM}(\omega)$ that are reliably encoded in the data do not suffer from the changes in $m_l$.

\subsection*{Acknowledgements}

The author would like to thank T. Hatsuda, O.Kaczmarek, J.-I. Skullerud and S. Sasaki for the many valuable discussions and comments. A.R. acknowledges support from the BMBF project \textit{Heavy Quarks as a Bridge between
Heavy Ion Collisions and QCD}, funding from the Sofja Kovalevskaja program of the Alexander von Humboldt foundation and the EU Integrated Infrastructure Initiative \textit{Hadron Physics 2} as well as partial support by the Swiss National Science Foundation
(SNF) under grant 200021-140234. 

\appendix

\section{Efficient marginalization of $\alpha$}

In this appendix I would like to mention a technical detail regarding the implementation of the procedure to marginalize the artificial parameter $\alpha$ inserted in Eq.\eqref{Eq:PriorProb}. To this end one calculates the maximum $\rho^\alpha$ of ${\cal Q}(\rho,D,m,\alpha)$ for many different values of $\alpha$ and then self consistently averages the results \cite{springerlink:10.1007/BF02427376,Jarrell1996133,Asakawa:2000tr,Nickel:2006mm} using the following relation
\begin{align}
 \rho^{\rm MEM}(\omega)&=\int {\cal D}\rho \int d\alpha \rho(\omega) P[\rho|D,I(m),\alpha]P[\alpha|D,I(m)]\\
&\simeq \int d\alpha \rho^{\alpha}(\omega)P[\alpha|D,I(m)].
\end{align}
The explicit expression of $P[\alpha|D,I(m)]$ has been shown to be
\begin{align}
 &P[\alpha|D,I(m)]\propto\\
&{\rm exp}\Big[ {\cal Q}[D,\rho^{\alpha},I(m)]+\frac{1}{2}\sum_{k=0}^{N_\tau-1} {\rm log}\Big(\frac{\alpha}{\alpha+\Delta \omega\lambda_k}\Big)\Big],
\end{align}
where the $\lambda_k$ are the $N_\tau$ non-zero eigenvalues of the matrix
\begin{align}
 \Lambda^\alpha_{ij}=\sqrt{\rho^\alpha_i}\left.\frac{\delta^2{\cal L}}{\delta \rho_i\delta\rho_j}\right|_{\rho=\rho^\alpha}\sqrt{\rho^\alpha_i}.
\end{align}

Let us see why this symmetric $N_\omega\times N_\omega$ matrix only contains such a small number of nonzero eigenvalues. Using the SVD of $K^t=\bar{U}\bar{\Sigma} \bar{V}^t$ ( $\bar{U}$ is the $N_\omega\times N_\tau$ sized matrix consisting of the first $N_\tau$ columns of the matrix $U$ in Eq. \eqref{Eq:DefSVDsp} and $\bar{\Sigma}$  and $\bar{V}$ the corresponding matrices of size $N_\tau\times N_\tau$) we can rewrite ($\sqrt{\rho^\alpha}$ denotes the vector obtained after applying the square root to each individual component $\rho^\alpha_l$) 
\begin{align}
 \Lambda^\alpha= \sqrt{\rho^\alpha} \bar{U} \bar{\Sigma} \bar{V}^t \left.\frac{\delta^2{\cal L}}{\delta D^\rho\delta D^\rho}\right|_{\rho=\rho^\alpha} \bar{V} \bar{\Sigma} \bar{U}^t \sqrt{\rho^\alpha}.
\end{align}
With the additional definition of the two symmetric $N_\tau\times N_\tau$ matrices
\begin{align}
 &M = \bar{\Sigma} \bar{V}^t \left.\frac{\delta^2{\cal L}}{\delta D^\rho\delta D^\rho}\right|_{\rho=\rho^\alpha}\bar{V} \bar{\Sigma},\\
 &T = \bar{U}^t {\rm diag}[\rho] \bar{U}
\end{align}
and an application of Sylvester's determinant theorem we can rewrite the Eigenvalue equation for the matrix $\Lambda$ as
\begin{align}
 0&={\rm det}\Big[ \Lambda -\lambda_k I_{N_\omega \times N_\omega}\Big]\\
  &={\rm det}\Big[ \sqrt{\rho^\alpha} \bar{U} \bar{\Sigma} \bar{V}^t \frac{\delta^2{\cal L}}{\delta D^\rho\delta D^\rho} \bar{V} \bar{\Sigma} \bar{U}^t \sqrt{\rho^\alpha} - \lambda_k  I_{N_\omega \times N_\omega}\Big]\\
  &={\rm det}\Big[ \bar{\Sigma} \bar{V}^t \frac{\delta^2{\cal L}}{\delta D^\rho\delta D^\rho} \bar{V} \bar{\Sigma} \bar{U}^t \sqrt{\rho^\alpha}\sqrt{\rho^\alpha} \bar{U} - \lambda_k  I_{N_\tau \times N_\tau}\Big]\\
  &={\rm det}\Big[ MT - \lambda_k  I_{N_\tau \times N_\tau}\Big].
\end{align}

Here it is important to realize that the product of two symmetric matrices is not necessarily symmetric, i.e. $MT\neq TM$, so that algorithms for Hermitian matrices cannot be used. Of course, since the spectrum of the original matrix $\Lambda$ is real, the matrix $MT$ does not harbor any complex eigenvalues. In addition we see that the matrix $\Lambda$ contains two factors of the kernel $K$, which, in the case of  a large dynamical range in $K$, requires arithmetic of twice the precision compared to the rest of the procedure to yield correct values.

The reader should also be aware that the above manipulations are independent from our choice of search space. The matrix $\Lambda$ always contains $N_\tau$ non-zero eigenvalues even if we choose a search space with a different dimensionality.

\bibliographystyle{elsarticle-num-names}
\bibliography{references.bib}

\end{document}